# Comparing the latest ranking techniques: pros and cons of flexible skylines, regret minimization and skyline ranking queries

Davide, D.F., Foini

Politecnico di Milano
Milan, Italy
davide.foini@mail.polimi.it

**Abstract**

Long-established ranking approaches, such as top-k and skyline queries, have been thoroughly discussed and their drawbacks are well acknowledged. New techniques have been developed in recent years that try to combine traditional ones to overcome their limitations. In this paper we focus our attention on some of them: flexible skylines, regret minimization and skyline ranking queries, because, while these new methods are promising and have shown interesting results, a comparison between them is still not available. After a short introduction of each approach, we discuss analogies and differences between them with the advantages and disadvantages of every technique debated.

***Keywords:*** regret minimization, flexible skylines, skyline ranking, comparison

## 1 Introduction

Nowadays ranking methods have established themselves as an important tool in multi-criteria data aggregation; it is enough to think about a list of products on an online marketplace: a customer can select his/her preferences based on multiple parameters (e.g., price, reviews, the popularity of the seller) and the order in which the products are shown is computed accordingly. To tackle this problem, a lot of methods have been developed: the historical ones are top-k queries and skyline queries.

Top-k queries compute the k most relevant objects in the data set, while skyline queries [3] return the set of non-dominated tuples, which means the ones that are not dominated by any other tuple. A tuple $t_1$ dominates another tuple $t_2$ if $t_1$ has a greater value than $t_2$ for at least one parameter and a greater or equal value for all the other parameters. Both the approaches have been deeply discussed and their performances evaluated [1, 2, 4]. Even if both represent an important tool, they present drawbacks. Skyline queries need to examine all the tuples before getting a result and the size of the skyline increases considerably with the growth of dimensionality. Top-k queries on the other hand require the user to provide his preferences and he/she may not be able to properly state the weight to assign to every attribute.

To overcome these limitations further research focused on producing new frameworks that combine some of the features of the classical approaches with other algorithms.

This paper aims to analyse and propose a comparison of three of these new techniques: flexible skylines, regret minimization and skyline ranking. For each approach we provide a general introduction, we describe some algorithms developed and after that we provide a comparison between the algorithms analysed. The reason for this selection is because, even though many papers are available on the subjects, a focus on the analogies and differences between the three techniques is still missing.

The structure of the paper is the following: in Sections 2,3 and 4 we provide a description of each method and some algorithms that implement it; in Section 5 we discuss the differences and analogies between the approaches and in Section 6 we conclude the discussion.

## 2 Regret Minimization Queries

Regret minimization queries have been introduced in recent years [6], they combine top-k and skyline queries returning a controllable number of tuples, as the former, but without asking the user for preferences, as the latter.

The core notion of this method is the *regret ratio*: as thoroughly described in [6], it measures the dissatisfaction of the user with the k tuples selected and it can span from 0 (maximum happiness) to 1 (minimum happiness). Other important concepts are the *utility function* that estimates the utility of every point in the data set and the *gain*, which is equal to the maximum utility obtained from a subset of points in the data set. The users' utility functions are unknown (feature inherited from skylines). The aim of this kind of query is therefore obviously to keep the *maximum regret ratio* as low as possible. The *maximum regret ratio* is the worst possible regret obtainable considering all the users and all the utility functions in the family of utility functions considered.

To compute this type of query many algorithms have been proposed and we summarize some of them.

### 2.1 Cube and Greedy [6]

These two algorithms have been proposed by Nanongkai *et al.* and they deal with linear utility functions, and both take as input a dataset of points of multiple dimensions.

Cube firstly computes the set of points with maximum value for each dimension (except for the last one), then it divides each dimension (except for the last one) in a chosen number of intervals with equal size and creates a bucket of points for each interval. Finally, it outputs the point with the highest value of the last coordinate of every bucket.

Greedy, as obvious from its name, uses a greedy procedure. It starts selecting the point with the best value for the first coordinate and then it starts iterating, adding for every iteration the point that currently maximizes the regret ratio to the result set.

### 2.2 UtilityApprox [7]

This algorithm is the result of further research following [6], in fact, it was developed with the enhancing of interaction: the user is allowed to state his preferred tuple among the ones proposed to him. The main idea behind this algorithm is to approximate the utility function of the user minimizing the difference between the one of the real user (denoted *u*) and the one of a virtual user (denoted *v*). The difference between the two utility functions is repeatedly reduced showing the user a set of points so that after one is chosen, the gap is reduced. The procedure is repeated for several rounds to reduce the difference between the utility functions by a significant factor.

While experiments show that the interactions increase the overall efficiency, they also introduce drawbacks: the improvement is not as relevant with correlated data and a small number of rounds is required to outdo the other algorithms that require just one iteration. It also deals with linear utility functions, such as Cube and Greedy.

## 2.3 Sorting-Simplex and Sorting-Random [8]

To mitigate the number of iterations requested by existing regret minimization methods as [7], these algorithms introduce interactions based on a sorting mechanism: the user is asked to rank the tuples that are proposed to him/her instead of choosing one of them. The sorting carried out by the user enables the algorithm to reduce the *utility space* (the set of all the possible utility functions) and so to get the maximum regret ratio more quickly. The difference between the two is the method of selection of the tuples to propose to the user: Sorting-Simplex selects the points using the concept of Simplex method for Linear Programming problem, while Sorting-Random selects the points randomly.

Both algorithms have shown better performances on synthetic and real datasets than other algorithms, among them UtilityApprox [7] (for further information please refer to [8 sections 4 and 5]).

## 2.4 2d kRMS [9]

The algorithm works in a two dimensions point space: each point $p(x_p; y_p)$ is transformed into a line with equation $x_p x + y_p y = 1$, orthogonal to the segment that connects the origin of the axis to point *p*. A *convex chain* is a sequence of consecutive line segments where each segment has a negative slope and it's shorter than the next one in the sequence. Via this transformation, it is possible to represent a set of points through a convex chain beginning on the positive x-axis and ending on the positive y-axis. Points belonging to the lines in the dual space form *top-k rank contours*: the set of points from each line such that for any point belonging to the contour, exactly *k-1* lines in the space cross the segment *O-p*. In Figure 1 and Figure 2 is given a graphical representation of the previous concepts.

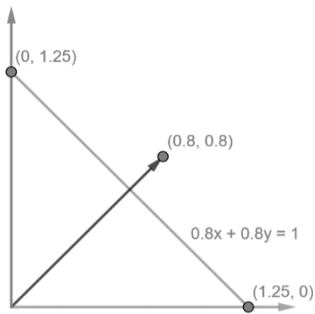

Figure 1: an example of transformation. The point (0.8, 0.8) is transformed into the orthogonal line 0.8x + 0.8y = 1

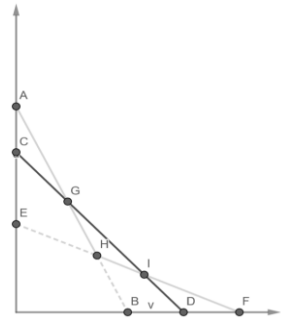

Figure 2: an example of convex chain and top-1 contour. An example of the former is ((D, G), (G, C)) and of the latter is (B, H, E)

Generally, the procedure consists of a radial plane sweep that keeps track of the convex chains detected. During the computation the best solution found so far is kept and the list is updated until the end of the rotation, which starts from the x-axis and terminates on the y-axis, when the final list is returned.

The experiments conducted have shown that values of k > 1 bring the best performances.

# 3 Flexible Skylines

Flexible skylines, also called restricted skylines, were first developed [5] to reconcile scoring functions, taken from classical top-k queries, and skyline queries. They alter the concept of *dominance* in different ways.

## 3.1 R-Skylines [5]

Restricted skylines (R-skylines) extend the concept of dominance to F-dominance: a tuple $t_1$ F-dominates another tuple $t_2$ when $t_1$ dominates $t_2$ according to all the scoring functions in the family of scoring functions F.

They introduced two new operators: *non-dominated restricted skyline* and *potentially optimal restricted skyline*. The *non-dominated restricted skyline*, referred to as ND, is composed of all the tuples that are not F-dominated, while the *potentially optimal restricted skyline*, referred to as PO, includes the top-1 tuples according to some scoring functions of F.

To compute ND the authors proposed different algorithms that combine different possibilities: whether to sort the data beforehand or not, whether to test the F-dominance solving an LP problem or checking if a tuple belongs to the *F-dominance region* of the tuples in the dataset and whether to compute ND after computing the skyline or do it in one phase. The *F-dominance region* of a tuple under a set of monotone scoring functions is the set of all the points that are F-dominated by the considered tuple.

To compute PO the tuples of ND are progressively discarded if they are F-dominated by a convex combination of tuples in ND and to do so in a more efficient way some heuristics are adopted.

Further research [10] (which renamed restricted skylines to flexible skylines, referred to as F-skylines) introduced more algorithms to compute PO based on different possibilities as for ND. The algorithms can be of one or two phases, use a primal or dual PO test and use a set of tuples with an incremental size, or use just the full-sized set. Therefore, there are eight alternatives for computing ND and eight for computing PO.

Concerning skyline queries, R-skylines have shown big improvements reducing the number of tuples returned, but concerning ranking queries, they have the drawback of higher computational time due to the advantage of returning the most interesting points.

## 3.2 RSA and JAA [11]

These two algorithms were developed with the concept of *uncertain top-k queries* (UTK): the reason for this is that the user preferences can be estimated only boundedly or they can be uncertain since the user can be incapable of selecting exact weights for his/her preferences. UTK can compute the tuples that could be part of the top-k set or report the exact top-k set for all the possible preferences settings: for the first option the algorithm proposed is RSA and for the second one the algorithm is JAA. Both algorithms assume that the dataset is structured by a spatial index like an R-tree [18].

Both use the notion of *r-dominance*, which is like F-dominance but uses weight vectors in R instead of scoring functions in F.

RSA (R-Skyband Algorithm) first filters the dataset eliminating the tuples that do not belong to the r-skyband, namely the set of tuples that are r-dominated by less than *k* others, and then builds the r-dominance graph: a DAG (Directed Acyclic Graph) in which an arc from a node $n_1$ to node $n_2$ implies that $n_1$ r-dominates $n_2$.

The next phase is the refinement phase, in which the candidates are examined in decreasing order of their r-dominance counts and selected as part of the result or not. The order is chosen to reduce the number of candidates to examine: if a candidate is confirmed to be part of the result set, its ancestor will be also added to the result, avoiding examining them.

JAA (Joint Arrangement Algorithm) differs from RSA in the refinement phase: it works on a common global arrangement for all the candidates instead of verifying all the candidates independently. The main concept is the *anchor record*, which is used for the verification of candidates as in RSA. The anchor is first selected among the tuples received from the filtering phase and different choosing strategies are proposed to update the anchor. JAA terminates only when every partition of R is classified as a less-than, a greater-than, or an equal-to partition. This classification is referred to as the rank of the anchor *k'* and the one of the verified partition *k*: in less-than partitions *k' < k*, in greater-than ones *k' > k* and *k' = k* for equal-to partitions.

The performances of the two algorithms have been compared to baseline algorithms obtained combining already existing procedures (for further information please refer to [11 section 3.3]) and they have both outperformed them by one to two orders of magnitude.

## 3.3 ORD and ORU [12]

These two algorithms were developed to overcome some limitations of [11] and [5]: they are not output size specified (OSS), meaning that the size of the result cannot be dictated, and they need the user to specify a region in the preference domain. ORD and ORU require a weight vector as input and expand it in different ways: the former uses an adaptive concept of dominance based on the output size (dominance-based) and the latter is similar to ranking by utility (utility-based).

A crucial notion is the one of ρ-dominance: considering the preference vector that estimates the user preferences $w$ (called *seed*) and all the other vectors $v$ within distance ρ from $w$, a record $r_1$ *ρ-dominates* $r_2$ if $r_1$ scores at least as good as $r_2$ for every vector $v$ and strictly better for at least one vector $v$.

Like RSA and JAA, described in the previous section, both ORD and ORU work with a dataset ordered by a spatial index.

ORD takes as input the weight vector and the desired output size $m$ and produces as output the set of records that are ρ-dominated by less than $k$ others, of the minimum ρ that produces exactly the desired output size, namely the ρ-skyband.

A simplified version of ORD:
1. Computes the full k-skyband.
2. Derives the *inflection radius* for each record in it. The *inflection radius* is the value of ρ for which the record becomes part of the ρ-skyband.
3. The $m$ records with the smallest inflection radii are returned.

ORU takes the same inputs as ORD, but it produces the records that belong to the top-k set for at least one weight vector within a certain distance from the input weight vector so that exactly the number of desired records is produced.

The core notions for ORU are the *convex hull* and the *hull layers*. The *convex hull* can be defined as the smallest convex polytope that contains all the records of a dataset and it is composed of facets that have two records as vertices. The *hull layers* are obtained recursively from the data set: the first layer is $L_1$ is the *upper hull* of D, the second layer $L_2$ is the *upper hull* of D − $L_1$, and so on, where the *upper hull* contains the facets defined by the extreme vertices with non-negative *norms* (the *norm* of a facet is the normal vector whose sum of coordinates is 1 and is directed towards the exterior of the hull) of the *convex hull*. In Figure 3 we provide an example of the previous concepts.

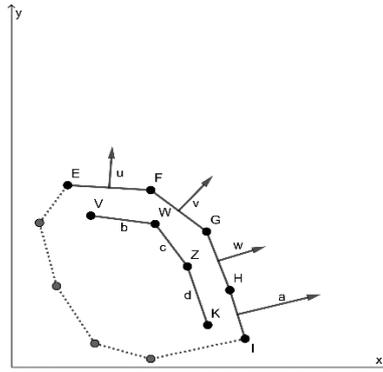

*Figure 3: an example of a convex hull and two layers. The convex hull is the outer polygon, while the first layer is composed of records from E to I and the second is composed of records from V to K*

Assuming that the minimum radius $ρ$ to produce the required number of records and the first $k$ upper hull layers are already available, ORU proceeds as follows:
1. The records in the first layer are filtered: only the ones that have their top-region within a maximum distance from the weight vector of $ρ$ are selected.
2. Every top-region is isolated and the top-2-nd record in it is found and the region is partitioned consequently.

3. The same process is executed recursively in each partition until the full and ordered top-k is known anywhere in the top-region and the same is done for every top-region of the records selected in the first layer.
4. The required top-k result is returned.

Both ORD and ORU showed better performances than previous techniques and demonstrated the ability to scale well with all the parameters.

# 4 Skyline Ranking

One of the drawbacks of skyline queries is that the points belonging to the result set are all equal: it is not possible to determine if a point is "better" than another. To overcome this drawback an automated ranking procedure has been introduced to return to the user just the top-k interesting points in the skyline set.

## 4.1 SKYRANK [13]

To be able to give a score to each skyline point, SKYRANK uses the notion of *subspace dominance relationship*, which limits the more general concept of dominance used in classical skylines to a subspace of the original data space. Therefore, a skyline point is considered more interesting if it dominates many other skyline points in different subspaces and this quality is transferred to all the points that dominate it in a subspace. Another essential tool is the *skyline graph*, which is a weighted directed graph in which:
- The vertices correspond to the skyline points.
- There is an edge between two points $p$ and $q$ if $p$ dominates $q$ in any subspace.
- The weight of an edge is computed as the ratio between the number of subspaces in which the subspace dominance between the two vertices exists and the total amount of subspaces in which the subspace dominance exists considering all the graph.

SKYRANK takes as input a *skyline graph* and applies a link-based ranking algorithm, based on PageRank [17], which values more the points that dominate many other points, that iteratively dominate many others in some subspaces. User preferences can also be considered expressing them as a set of preferred subspaces and by favouring skyline points that belong to the preferred subspaces without rebuilding the *skyline graph*.

To evaluate SKYRANK, it has been compared with two other approaches: the *skyline frequency metric*, which measures the frequency of a point considering different subspaces, and the *subspace dominance metric*, which considers the number of points that a skyline point dominates. The experiments carried out have shown that SKYRANK is a robust algorithm, and it provides a meaningful ranking.

## 4.2 SCSQuerySkyOrdPre [14]

This algorithm introduces the notion of *skyline order*: a set of $n$ skylines for a set of d-dimensional points in which the first skyline is the one obtained considering all the points and all the other ones are the skylines of the dataset without the points belonging to the preceding skylines in the order. The union of all the skylines in the ordered set is equal to the original data set and the intersection between them is therefore empty. Each skyline in the ordered set is called *skyline subset* or *skyline order subset*.

Before computing the *skyline order,* the data set is sorted in nondecreasing order of the sum of all the dimensional values of each point, so to avoid comparing a point with the ones computed before it because it cannot dominate them. The computation is performed via a loop for every point $p$ in the dataset:
- For every *skyline order set*: if $p$ is not dominated by any point in the ordered set, add it to the ordered set.
- If $p$ was not inserted in any subset, create a new subset with just $p$ and add it to the sets list.

SCSQuerySkyOrdPre (abbreviation for Size-Constrained Skyline Query with Skyline Ordering Precomputed) takes as input a *skyline order* and the number of points to retrieve $k$. It loops on the precomputed skyline order subsets maintaining a counter of the points analysed. Starting from an empty result set, for each order subset:
- If its cardinality is equal to *k*, it is merged to the result set and the algorithm stops.
- Otherwise, if its cardinality is lower than *k*, it is merged to the result set, and *k* is lowered by its cardinality.
- Otherwise, *k* points are selected from the ordered subset, added to the result and the algorithm stops (the method to do so is not reported, for further information please refer to [14 section 3.2]).

Another version of SCSQuerySkyOrdPre is SCSQuerySkyOrd, which does not need a precomputed skyline order, but just a sorted data set. The difference from the previous algorithm is that it only keeps a partial skyline order and as soon as there are enough points SCSQuerySkyOrdPre is called with the partial skyline order as input, saving substantial computational cost.

### 4.3 ZRank [15]

The essential concepts to understand how ZRank works are:
- *ZBtree,* an indexing structure for skyline candidates.
- *Z-order curve,* a Z-order space-filling curve.
- *RZ-region*, the smallest square region covering a bounded Z-region

The algorithm is composed of two phases: the *counting phase*, which scans the skyline candidates counting how many points they dominate, and the *ranking phase*, which orders the candidates on their dominating power, returning the first *t* elements, where *t* is an input parameter.

ZRank takes as input a ZBtree that indexes the skyline points and the desired number of tuples to return, then outputs the top-ranked skyline points and keeps as local variables a stack and a new ZBtree.
It examines child nodes of the non-leaf nodes and if they need to be explored, they are added to the stack. For leaf nodes, the child points are compared to the ones in the local tree and if they are not dominated, they are inserted in it. After all the entries are examined the candidates in the local tree are ranked and the first selected number of them is returned.

ZRank has been compared to two SFSRank and BBSRank [16] and it clearly outperformed them in the experiments realized.

## 5 Comparison

In this section, we provide a comparison between the algorithms analysed of the three considered methods on different aspects: the control of the output size, the need of user interaction, the usage of internal parameters, the usage of indexes and the type of ranking function used.

### 5.1 Output Size Control

The first term of comparison is the feature of controlling the output size passing the desired one as input to the algorithm. This characteristic is not homogeneous for each type of approach: independently from it, some among the analysed algorithms have this feature, and others do not have it. Cube, Greedy, UtilityApprox, 2d kRMS, RSA, JAA, ORD, ORU, SCSQuerySkyOrdPre, and ZRank return the desired number of tuples as a result while the others cannot.

### 5.2 User Interaction

This feature is instead homogeneous for the algorithms of each type: the user interaction is required for the algorithms that compute regret minimization queries, while the ones that compute flexible skylines queries and skyline ordering queries do not have that as a feature.

## 5.3 Parameters

Another term of comparison is the need for internal parameters to be calibrated and that can affect the result obtained. UtilityApprox uses some parameters: the first is $s$, the number of tuples to show at each round, and other parameters are specified in the sub-algorithm that chooses the points to display. ORD and ORU have the parameter $k$, which for ORD represents the maximum number of tuples that can ρ-dominate a tuple in the result set, and in ORU is used to compute the top-$k$ sets for every weight vector. All the other algorithms don't need internal parameters.

## 5.4 Indexes required

Among the algorithms analysed the only one that requires indexes are RSA and JAA, which require a spatial index, ORD and ORU, which also require a spatial index, and ZRank that requires an index for the ZBtree.

## 5.5 Ranking Function

The ranking function, also referred to as scoring function or evaluation function, can belong to different families of functions. All the algorithms considered that compute regret minimization queries deal with linear ranking functions, while for flexible skylines queries restricted skylines deal with monotone ranking functions, RSA and JAA with linear functions for the preference weights, and monotone ones for the score records and ORD and ORU with linear evaluation functions. The algorithms that compute skyline ranking queries for their nature deal with generic types of scoring functions.

## 5.6 Summary

In the following table, we report a summary of the comparison given in the previous paragraphs.

| Algorithm | Output Size Controlled | User Interaction | Parameters Free | Indexes required | Ranking Function | | |
|---|---|---|---|---|---|---|---|
| | | | | | Linear | Monotone | Generic |
| *Regret Minimization* | | | | | | | |
| Cube and Greedy [6] | ✓ | ✓ | ✓ | ✗ | ✓ | | |
| UtilityApprox [7] | ✓ | ✓ | ✗ | ✗ | ✓ | | |
| Sorting-Simplex and Sorting-Random [8] | ✗ | ✓ | ✓ | ✗ | ✓ | | |
| 2d kRMS [9] | ✓ | ✓ | ✓ | ✗ | ✓ | | |
| *Flexible Skylines* | | | | | | | |
| R-Skylines [5] | ✗ | ✗ | ✓ | ✗ | | ✓ | |
| RSA and JAA [11] | ✓ | ✗ | ✓ | ✓ | ✓ | ✓ | |
| ORD and ORU [12] | ✓ | ✗ | ✗ | ✓ | ✓ | | |
| *Skyline Ranking* | | | | | | | |
| SKYRANK [13] | ✗ | ✗ | ✓ | ✗ | | | ✓ |
| SCSQuerySkyOrdPre [14] | ✓ | ✗ | ✓ | ✗ | | | ✓ |
| ZRank [15] | ✓ | ✗ | ✓ | ✓ | | | ✓ |

*Table 1: summary of the comparison*

# 6 Conclusion

In this paper, we provided a discussion about some algorithms that implement the latest ranking techniques, namely regret minimization queries, flexible skyline queries, and skyline ranking queries. After that, we compared the algorithms on different aspects and summarized the comparison in a table. Further research could focus on testing the algorithms on different datasets to effectively compare them in terms of computation time and the ranking produced as output.

# References

[1] Ilyas, I. F., Beskales, G., & Soliman, M. A. (2008). A survey of top-k query processing techniques in relational database systems. *ACM Computing Surveys*, 40(4).

[2] Bruno, N., Chaudhuri, S., & Gravano, L. (2002). Top-k Selection Queries over Relational Databases: Mapping Strategies and Performance Evaluation. *ACM Transactions on Database Systems*, 27(2), 153–187.

[3] Börzsönyi, S., Kossmann, D., & Stocker, K. (2001). The skyline operator. *Proceedings - International Conference on Data Engineering*, 421–430.

[4] Chomicki, J., Ciaccia, P., & Meneghetti, N. (2013). Skyline queries, front and back. *SIGMOD Record*, 42(3), 6–18.

[5] Ciaccia, P., & Martinenghi, D. (2017). Reconciling skyline and ranking queries. *Proceedings of the VLDB Endowment*, 10(11), 1454–1465.

[6] Nanongkai, D., Sarma, A. Das, Lall, A., Lipton, R. J., & Xu, J. (2010). Regret-minimizing representative databases. *Proceedings of the VLDB Endowment*, 3(1), 1114–1124.

[7] Nanongkai, D., Lall, A., Das Sarma, A., & Makino, K. (2012). Interactive regret minimization. *Proceedings of the ACM SIGMOD International Conference on Management of Data*, 109–120.

[8] Zheng, J., & Chen, C. (2020). Sorting-Based Interactive Regret Minimization. *APWeb/WAIM (2) 2020*: 473-490

[9] Chester, S., Thomo, A., Venkatesh, S., & Whitesides, S. (2014). Computing k-regret minimizing sets. *Proceedings of the VLDB Endowment*, 7(5), 389–400.

[10] Ciaccia, P., & Martinenghi, D. (2020). Flexible Skylines: Dominance for Arbitrary Sets of Monotone Functions. *ACM Transactions on Database Systems*, 45(4), 1–45.

[11] Mouratidis, K., & Tang, B. (2018). Exact processing of uncertain top-k queries in multicriteria settings. *Proceedings of the VLDB Endowment*, 11(8), 866–879.

[12] Mouratidis, K., Li, K., & Tang, B. (2021). Marrying Top-k with Skyline Queries: Relaxing the Preference Input while Producing Output of Controllable Size. *Proceedings of the ACM SIGMOD International Conference on Management of Data*, 1317–1330.

[13] Vlachou, A., & Vazirgiannis, M. (2010). Ranking the sky: Discovering the importance of skyline points through subspace dominance relationships. *Data and Knowledge Engineering*, 69(9), 943–964.

[14] Lu, H., Jensen, C. S., & Zhang, Z. (2011). Flexible and efficient resolution of skyline query size constraints. *IEEE Transactions on Knowledge and Data Engineering*, 23(7), 991–1005.

[15] Lee, K. C. K., Lee, W. C., Zheng, B., Li, H., & Tian, Y. (2010). Z-SKY: An efficient skyline query processing framework based on Z-order. *VLDB Journal*, 19(3), 333–362.

[16] Chan, C. Y., Jagadish, H. V., Tan, K. L., Tung, A. K. H., & Zhang, Z. (2006). Finding k-dominant skylines in high dimensional space. *Proceedings of the ACM SIGMOD International Conference on Management of Data*, 503–514.

[17] Brin, S., & Page, L. (1998). The anatomy of a large-scale hypertextual Web search engine BT - Computer Networks and ISDN Systems. *Computer Networks and ISDN Systems*, 30(1–7), 107–117.

[18] Guttman, A. (1984). R-trees: A dynamic index structure for spatial searching. *ACM SIGMOD Record*, 14(2), 47–57.